\PassOptionsToPackage{table,dvipsnames}{xcolor}

\documentclass[sigconf,screen]{acmart}
\AtBeginDocument{%
  }

\setcopyright{cc}
\setcctype{by}
\acmDOI{10.1145/3832783.3837534}
\acmYear{2026}
\copyrightyear{2026}
\acmISBN{979-8-4007-2882-2/2026/10}
\acmConference[ASE '26]{Proceedings of the 41st IEEE/ACM International Conference on Automated Software Engineering}{October 12--16, 2026}{Munich, Germany}
\acmBooktitle{Proceedings of the 41st IEEE/ACM International Conference on Automated Software Engineering (ASE '26), October 12--16, 2026, Munich, Germany}
\acmSubmissionID{ase26main-p2532-p}
\received{2026-03-26}
\received[accepted]{2026-06-18}

\usepackage{booktabs}
\usepackage{tabularx}
\usepackage{multirow}
\usepackage{amsmath}
\usepackage{enumitem}
\usepackage{graphicx}
\usepackage{textcomp}
\usepackage{xcolor}
\usepackage{makecell}
\usepackage{subcaption}
\usepackage{listings}
\usepackage{fancyhdr}
\usepackage{arydshln}
\usepackage{multirow}
\usepackage{soul}
\usepackage{colortbl}  
\definecolor{probeorange}{RGB}{255, 247, 235}

\usepackage{algorithm}
\usepackage{algpseudocode}
\definecolor{mygreen}{HTML}{2DA44E}
\usepackage{balance}
\usepackage[most]{tcolorbox}
\usepackage{subcaption}
\tcbuselibrary{skins,breakable}

\newcommand{\phead}[1]{\vspace{1mm} \noindent {\bf #1}}

\definecolor{findingsback}{RGB}{230,247,240}
\definecolor{findingsborder}{RGB}{24,165,123}
\newtcolorbox{findings}{
    enhanced,
    breakable,
    colback=findingsback,
    colframe=white,
    borderline west={3pt}{0pt}{findingsborder},
    boxrule=0pt,
    arc=0pt,
    sharp corners,
    left=2mm, right=2mm, top=1mm, bottom=1mm,
}
\usepackage{url}
\usepackage{xspace}
\newcommand{\tool}{\textit{PROBE}\xspace}

  \usepackage{listings}
  \usepackage{subcaption}
  \usepackage{xcolor}

\definecolor{diffgreen}{HTML}{006400}
\definecolor{diffgray}{HTML}{666666}
\lstdefinestyle{codestyle}{
  basicstyle=\ttfamily\fontsize{7}{8.5}\selectfont,
  breaklines=true,
  breakatwhitespace=false,
  columns=fullflexible,
  keepspaces=true,
  xleftmargin=2pt,
  xrightmargin=2pt,
  aboveskip=3pt,
  belowskip=3pt,
  frame=none,
}
\lstdefinestyle{tinycode}{
  basicstyle=\ttfamily\fontsize{6}{7.2}\selectfont,
  breaklines=true, columns=fullflexible,
  keepspaces=true, xleftmargin=1pt,
  xrightmargin=1pt, aboveskip=1pt,
  belowskip=1pt, frame=none,
}

\definecolor{diffadd}{HTML}{DAFBE1}
\definecolor{diffaddtext}{HTML}{1A7F37}
\definecolor{diffdel}{HTML}{FFCECB}
\definecolor{diffdeltext}{HTML}{CF222E}
\definecolor{codebg}{HTML}{F6F8FA}
\definecolor{framegray}{HTML}{D0D7DE}
\definecolor{codegray}{HTML}{636C76}
\definecolor{warnbg}{HTML}{FFF8C5}
\definecolor{warntext}{HTML}{9A6700}

\lstdefinelanguage{diff}{
  morecomment=[f][\color{diffdeltext}]-,
  morecomment=[f][\color{diffaddtext}]+,
  morecomment=[f][\color{codegray}]\#,
}

\lstdefinestyle{diffstyle}{
  language=diff,
  basicstyle=\ttfamily\small,
  columns=fullflexible,
  keepspaces=true,
  showstringspaces=false,
  aboveskip=0pt,
  belowskip=0pt,
  moredelim=[is][\color{diffaddtext}\begingroup\let\lst@linebg\@gobble]{+BGadd}{+BGadd},
}

\newtcolorbox{diffbox}{
  enhanced, colback=codebg, colframe=framegray,
  boxrule=0.5pt, arc=1.5pt,
  left=5pt, right=5pt, top=4pt, bottom=4pt,
}
\usepackage{etoolbox}
\makeatletter
\patchcmd{\@mkbibcitation}{\par\nobreak}{\par}{}{\typeout{PATCHFAIL}}
\pretocmd{\@mkbibcitation}{\clubpenalty=0 \widowpenalty=0 \interlinepenalty=0 \brokenpenalty=0 }{}{\typeout{PRETOFAIL}}
\makeatother
\begin{document}

\title{Probe to Generate: Program Variant-Guided Test Augmentation for Repository-Level Repair Benchmarks}

\author{Chenglin Li}
\authornote{Equal contribution.}
\email{chenglin.li@mail.concordia.ca}
\affiliation{%
  \institution{SPEAR lab, Concordia University}
  \city{Montreal}
  \state{Quebec}
  \country{Canada}
}

\author{Yisen Xu}
\email{yisen.xu@mail.concordia.ca}
\authornotemark[1]
\affiliation{%
  \institution{SPEAR lab, Concordia University}
  \city{Montreal}
  \state{Quebec}
  \country{Canada}
}
\author{Zehao Wang}
\email{w\_zeha@encs.concordia.ca}
\affiliation{%
  \institution{SPEAR lab, Concordia University}
  \city{Montreal}
  \state{Quebec}
  \country{Canada}
}

\author{Shin Hwei Tan}
\email{shinhwei.tan@concordia.ca}
\affiliation{%
  \institution{Concordia University}
  \city{Montreal}
  \state{Quebec}
  \country{Canada}
}
\author{Tse-Hsun (Peter) Chen}
\email{peterc@encs.concordia.ca}
\affiliation{%
  \institution{SPEAR lab, Concordia University}
  \city{Montreal}
  \state{Quebec}
  \country{Canada}
}


\begin{abstract}
Test-based benchmarks such as SWE-bench have become a standard basis for evaluating automated issue resolution agents, deeming a patch correct if it passes a provided regression test suite. In practice, weak test suites can admit plausible but semantically incorrect patches, inflating reported agent performance. We present \tool, a test augmentation framework that uses semantically modified program variants as behavioral probes to identify and close gaps in benchmark test suites. Variants of the reference patch that survive the original tests reveal under-constrained behaviors, which then guide targeted regression test generation. Each generated test is retained only if it passes on the reference patch, fails on at least one surviving variant, and remains robust under behavior-preserving transformations. On SWE-bench Verified, 77\% of instances admit at least one surviving variant. \tool generates 1,014 validated tests across 211 instances, increasing patch-region line and branch coverage by 10.8 and 9.5 percentage points. Re-evaluating the top-10 repair agents with the augmented suites reduces resolved rates by 4.2\%--9.0\%, showing that many previously accepted patches exploit benchmark test gaps rather than fully satisfying the intended repair semantics. These findings demonstrate that benchmark evaluation is not solely a patch-generation problem but also a test-strength problem.
\end{abstract}




\maketitle

\section{Introduction}  

Recent advances in large language models (LLMs) have enabled significant progress in automated issue resolution agents~\cite{xia2025live,Sonar,gao2025trae,Rovo,epam,acoder,Warp,harness}, particularly on benchmarks such as SWE-bench~\cite{jimenez2023swe}. These benchmarks evaluate agents by executing generated patches against a provided regression test suite and deeming a patch correct if it passes all tests. This evaluation paradigm has become a common practice due to its scalability and reproducibility, and it underpins most recent claims of progress in issue resolution.

However, this evaluation paradigm relies on an underlying assumption: that the regression test suite provides a sufficiently complete specification of the intended behavior. In practice, this assumption rarely holds. Regression tests in issue-driven benchmarks are typically designed to validate observed symptoms described in issue reports, rather than to comprehensively encode the full behavioral intent of the fix. As a result, patches that are \emph{plausible but semantically incorrect} can still pass all tests. This phenomenon, commonly referred to as \emph{test-suite overfitting}~\cite{qi2015analysis,smith2015cure}, leads to systematic overestimation of agent performance.

Recent studies~\cite{wang2025solved, aleithan2024swe, yu2025utboost} show that even carefully curated benchmarks such as SWE-bench Verified~\cite{swebench_verified} admit a non-trivial number of behaviorally incorrect patches that nonetheless pass evaluation. Consequently, current benchmarks risk conflating \emph{test adequacy} with \emph{semantic correctness}, providing inaccurate assessment of the true capabilities and hiding the limitations of repair agents. This threatens the validity of empirical comparisons and may misguide future research directions.

Despite growing recognition of this problem, existing efforts exhibit three 
key limitations. First, prior studies~\cite{aleithan2024swe} rely on manual patch analysis, 
detecting incorrect fixes \emph{after} evaluation rather than strengthening 
the evaluation process itself. Second, test generation approaches for 
SWE-bench~\cite{wang2025solved} are designed to distinguish AI-generated patches from 
human-written ones---an objective misaligned with improving test suite 
adequacy. Third, test-augmentation approaches such as UTBoost~\cite{yu2025utboost} decouple 
identification from generation, foregoing diagnosis as a means to target 
specific gaps. This separation risks producing tests that redundantly exercise 
well-covered behavior while leaving under-specified behaviors unaddressed. 
Consequently, 
generated tests risk reinforcing coverage of already well-tested 
functionality while leaving critical behavioral gaps unaddressed. 
Bridging test inadequacy analysis with targeted test generation in a  unified framework remains an open problem.

In this paper, we argue that evaluating issue-resolution agents requires moving beyond passive reliance on existing regression tests toward \emph{adequacy-driven test augmentation}. We propose \tool, a fully automated framework that strengthens repair benchmarks by using program variants as \emph{behavioral probes}. Unlike mutation-guided test generation~\cite{mutgen}, which uses mutants as targets for improving mutation score, \tool uses variants to test whether the benchmark can distinguish the gold repair from plausible incorrect repairs. Specifically, \tool perturbs the gold patch to construct plausible incorrect repair variants and evaluates them using the benchmark's existing test suite. A variant that passes all existing tests exposes a concrete adequacy gap where the benchmark does not distinguish the gold repair from that of an incorrect alternative. \tool then generates a targeted contrastive test that passes on the gold repair but fails on the surviving variant, directly eliminating the ambiguity revealed by the probe.

PROBE combines two complementary strategies
to generate such variants: 1) operator-based mutation that produces
fine-grained and localized changes, and 2) LLM-based mutation that
introduces context-dependent modifications that fixed operators
are harder to reach. Guided by the surviving variants, PROBE then
generates targeted tests designed to distinguish correct implementations from plausible but incorrect ones. These surviving variants are then used to guide the generation of targeted tests that distinguish correct implementations from plausible but incorrect ones.  
Before incorporating into augmented test suite, each generated test must satisfy three
conditions: (i) it passes on the reference patch, (ii) fails on at least one
surviving variant, and (iii) avoids overfitting to implementation-specific
details, verified through behavior-preserving transformations and LLM-based screening. 
By tightly coupling \emph{diagnosis} and \emph{augmentation}, \textsc{PROBE} directly addresses the root cause of unreliable evaluation.

We conduct a study to evaluate \textsc{PROBE} on SWE-bench Verified, focusing on both test suite adequacy and its impact on agent evaluation. 
Our results revealed the prevalence of under-constrained test suites, with many benchmark instances accepting incorrect yet test-passing variants. \textsc{PROBE} help strengthen test suites by introducing targeted behavioral checks, leading to measurable reductions in reported success rates and changes in the relative ranking of state-of-the-art agents. These findings highlight that improving evaluation is essential to draw reliable conclusions about repair capabilities.

This paper makes the following contributions:

\noindent\textit{\textbf{Approach.}} 
We propose \tool, a program variant-guided framework that identifies behavioral gaps via surviving variants and
generates targeted tests to close those gaps.

\noindent \textit{\textbf{Empirical Study}}. We conduct a large-scale evaluation on SWE-bench Verified, providing quantitative evidence of pervasive test suite inadequacy and its impact on evaluation reliability. Our study shows that 77\% of instances admit at least one surviving program variant, and identifying four common patterns in how existing tests fail to constrain the intended fix. These patterns include \emph{Insufficient Input Space Exploration}, \emph{Partial Patch Path Coverage}, \emph{Weak Assertions}, \emph{Missing Environmental Context}. 

\noindent\textit{\textbf{Impact on Evaluation.}} We show that strengthening test suites with \textsc{PROBE} leads to more discriminative evaluation (the augmented tests generated by
PROBE reduce the resolved rates of the top-10 repair agents
by 4.2\% to 9.0\% on SWE-bench Verified, about 2.75$\times$ more than the baseline on average), and shift the relative ranking of agents on the leaderboard. 

\noindent\textit{\textbf{Insights. }}We analyze the types of behavioral checks introduced by generated tests, offering insights into how targeted augmentation improves the detection of incorrect patches.

\phead{Paper Organization.}The rest of this paper is organized as follows: Section~\ref{sec:background} introduces the background and motivating example, and discusses related work. Section~\ref{sec:approach} presents the \tool framework. Section~\ref{sec:empirical} evaluates \tool through four research questions. Section~\ref{sec:threats} addresses threats to validity, and Section~\ref{sec:conclusion} concludes.
\section{Background and Related Work}\label{sec:background}

\subsection{Background}
We study this problem in the context of SWE-bench~\cite{jimenez2023swe}, a repository-level benchmark in which each instance pairs an issue description with a reference patch and regression tests drawn from the corresponding pull request, where the patch is the developer-written fix validated as correct and the tests pass on it. Benchmark construction pipelines generally adopt these developer-submitted tests as-is, without additional review or augmentation of their behavioral coverage. Recent studies have found that a substantial fraction of patches accepted on SWE-bench Verified are behaviorally incorrect~\cite{wang2025solved,aleithan2024swe,yu2025utboost}, confirming that this lack of test scrutiny has practical consequences. This raises our central question: \textbf{\emph{To what extent do SWE-bench regression test suites behaviorally constrain intended fixes, and can we systematically strengthen them?}}

Figure~\ref{fig:motivated} illustrates this problem with a real example from SWE-bench Verified (\texttt{django-11276}~\cite{django-11276}). The issue modifies how \texttt{escape()} formats its output. Although this change targets a single function, the new output format also affects \texttt{urlize()}, a downstream function that consumes and reverses the escaped text before processing URLs. The oracle patch updates both functions to keep them consistent. In contrast, the patch submitted by Trae~\cite{gao2025trae} updates only \texttt{escape()}, leaving \texttt{urlize()} expecting the old format. This patch passes all existing tests because the developer tests check only the local escape output---no test exercises the downstream URL-processing path.
 
This example highlights a recurring test weakness: existing tests may accept patches that fix the visible symptom while omitting semantically necessary downstream changes. \tool addresses this gap in two steps. 
First, \tool generates a surviving variant by making a small, behavior-changing edit inside the reference patch and keeping it only if it still passes all original tests. Here, the edit rewrites the downstream \texttt{urlize()} logic. As the developer-written test checks only the output of \texttt{escape()} and never exercises \texttt{urlize()}, the variant still passes all original tests, confirming that the downstream path is unconstrained. Second, \tool uses the surviving variant as a signal of this gap and prompts an LLM to generate a set of tests that the gold patch passes but the variant fails. Specifically, this test runs \texttt{escape()} and \texttt{urlize()} end-to-end and asserts on the final URL. It also fails on the plausible patch shown in Figure~\ref{fig:motivated}, exposing the missing downstream update.
By automatically generating targeted tests from surviving variants, \tool provides a practical, fully automated mechanism for improving benchmark reliability without requiring manual test authoring or domain-specific knowledge. 

\begin{figure}[t]
\small
\raggedright




\noindent\textbf{(a.) Oracle patch: updates both functions}

\vspace{2pt}
\begin{flushleft}
\ttfamily\scriptsize
def escape(text): ... \textcolor{green!50!black}{\# developer patch modified it to change text's format}\\

def urlize(text): ... \textcolor{green!50!black}{\# developer patch modified it to reverse the format change}
\\
\end{flushleft}

\vspace{6pt}

\noindent\textbf{(b). Plausible patch (by Trae): patch provided only for escape()}

\vspace{2pt}
\begin{flushleft}
\ttfamily\scriptsize
def escape(text):\quad\textcolor{green!50!black}{\checkmark\enspace patch provided}\\
def urlize(text):\quad\textcolor{red}{\texttimes\enspace NO patch provided}\\
\end{flushleft}

\vspace{6pt}

\noindent\textbf{(c). Developer test: checks escape() output only}

\vspace{2pt}
\begin{flushleft}
\ttfamily\scriptsize
assert escape(input) == expected\_entity\quad\textcolor{green!50!black}{\checkmark\enspace passes on both patches}\\
\end{flushleft}

\vspace{6pt}

\noindent\textbf{(d). Generated test (by \tool): checks both functions}

\vspace{2pt}
\begin{flushleft}
\ttfamily\scriptsize
assert escape(input) == expected\_entity\quad\textcolor{green!50!black}{\checkmark\enspace passes on both patches}\\
assert urlize(url\_with\_entity) == expected\_url\\
\quad\textcolor{green!50!black}{\checkmark\enspace passes on oracle patch}
\quad\textcolor{red}{\texttimes\enspace fails on Trae's plausible patch}
\end{flushleft}

\vspace{-1em}
\caption{Motivating example (simplified \texttt{django-11276} for easier understanding). The output of \texttt{escape()} is consumed by \texttt{urlize()}. The oracle patch updates both functions, but the plausible patch modifies only \texttt{escape()}. The developer test checks only the local output and passes on both patches, missing the incomplete fix. \tool’s generated test instead exercises the end-to-end pipeline and reveals the missing update in \texttt{urlize()}.} 
\label{fig:motivated}
\vspace{-1em}
\end{figure}

\subsection{Related Work}\label{sec:related}








\phead{Test Suite Weakness in Coding Benchmarks.}
The reliability of benchmark-based evaluation is bounded by the strength of its test oracles, a problem first characterized as test-suite overfitting in the APR literature~\cite{qi2015analysis,smith2015cure}, and several countermeasures have been proposed, including anti-pattern filtering~\cite{tan2016anti} and differential test generation~\cite{1identifyovefittiedpatches,Opad}. Recent studies confirm that this risk persists at scale. On SWE-bench Verified, 29.6\% of plausible patches are behaviorally incorrect~\cite{wang2025solved}, and resolution rates drop significantly after filtering weak tests~\cite{aleithan2024swe,yu2025utboost}.  Beyond issue benchmarks, EvalPlus~\cite{liu2023your} reveals 15--20\% pass-rate drops on HumanEval~\cite{chen2021evaluating} and MBPP~\cite{austin2021program} under strengthened tests. These findings establish that weak test oracles are a systemic bottleneck across benchmark paradigms, motivating automated approaches to strengthen them.

\phead{Mutation Testing.}
Mutation testing assesses test suite adequacy by injecting small syntactic faults and checking whether tests detect them~\cite{jia2010analysis,ojdanic2023mutation}. Classical operator-based tools such as Mutpy~\cite{mutpy}, Mutmut~\cite{mutmut}, and Cosmic Ray~\cite{cosmicray} apply predefined operators to individual constructs, while higher-order mutation composes multiple operators to approximate real faults~\cite{langdon2010efficient}. More recently, LLM-based approaches such as $\mu$BERT~\cite{bert} and LLMorpheus~\cite{tip2025llmorpheus} generate context-sensitive, naturalistic mutants that extend the behavioral space beyond what fixed operators can reach. 
Mutation-guided test generation, such as MUTGEN~\cite{mutgen}, uses mutants as optimization targets: an LLM is repeatedly given surviving mutants and asked to generate tests that kill them, with the goal of improving mutation score for a program. \tool uses mutants for a different purpose. Its goal is not to maximize mutation score, but to assess and strengthen the discriminative power of a repair benchmark. In \tool, a surviving variant is treated as evidence that the benchmark test suite cannot distinguish the gold repair from a plausible incorrect repair. The variant therefore serves as a diagnostic probe for a benchmark gap. Given such a gap, \tool generates a targeted contrastive test that must fail on the surviving variant while passing on the gold patch. Thus, whereas MUTGEN uses mutants as feedback for iterative test generation, \tool uses surviving repair variants as counterexamples that reveal under-specified benchmark behavior. Additionally, MUTGEN uses PITest’s conventional operators and coverage feedback, whereas PROBE combines predefined operators with LLM-based semantic mutation. Finally, \tool operates at the repository-level repair setting, where tests must validate real patches in their project context, rather than at the function-level setting targeted by prior mutation-guided test generation.

\phead{Test Augmentation and Oracle Improvement.}
Approaches to strengthening test suites can be broadly grouped by their generation strategy. Search-based tools, such as EvoSuite~\cite{fraser2011evosuite} and Pynguin~\cite{lukasczyk2022pynguin}, optimize for structural coverage criteria. LLM-based approaches generate semantically richer tests by conditioning on code context and natural-language specifications~\cite{lemieux2023codamosa,yuan2024evaluating}. In the APR context, Xin and Reiss~\cite{1identifyovefittiedpatches} generate tests via symbolic execution to distinguish correct patches from overfitted ones, and Yang et al.~\cite{Opad} use specification-guided generation for patch assessment. For oracle quality specifically, Jahangirova et al.~\cite{jahangirova2016test} use mutation testing to identify false positives and false negatives in existing oracles, and Xie~\cite{xie2006augmenting} augments test suites with regression oracle checking by capturing object states and asserting on observer methods.  EvalPlus~\cite{liu2023your} augments function-level benchmarks with LLM- and mutation-based test input generation. Other techniques instead assess overfitting on a single given patch ~\cite{invalidator,Shibboleth,Fixchecker}, whereas PROBE strengthens the benchmark's tests so that they constrain any future patch.

Three recent works are most closely related to \tool. Wang et al.~\cite{wang2025solved} identify behaviorally incorrect patches on SWE-bench Verified through retrospective empirical analysis, but do not generate stronger tests to prevent such patches from being accepted. Meta's ACH~\cite{harman2025mutation} combines mutation and LLM-based test generation, but targets production-level regression hardening where fault classes are provided by engineers rather than discovered automatically. UTBoost~\cite{yu2025utboost} is most directly comparable, generating tests from code context and issue descriptions to strengthen SWE-bench evaluation. Unlike \tool, however, its weakness detection and test validation are both coupled to the availability of agent-generated patches, making adequacy assessment dependent on the particular agents under evaluation rather than an independent property of the benchmark.


All prior approaches share an important intuition: additional tests can reveal errors missed by the original suite. \textit{\textbf{What they do not address is the question of where the existing test suite is behaviorally insufficient, which semantic regions remain under-constrained and would most benefit from additional testing}}. Without this diagnostic step, generated tests may redundantly exercise already well-tested behaviors while leaving critical gaps untouched. PROBE addresses this by diagnosing these gaps before generating tests to close them. It first uses program variant generation to locate under-constrained behaviors, then uses the identified gaps, implemented as surviving variants, as contrastive signals for targeted test synthesis. Finally, we validate generated tests through behavior-preserving transformations to guard against overfitting to the oracle patch. This end-to-end process, from diagnosis to generation to validation, distinguishes PROBE from prior work that treats test generation and oracle assessment as separate concerns. 

\section{Approach}\label{sec:approach}

\begin{figure*}[t]
  \centering
\includegraphics[width=0.9\linewidth]{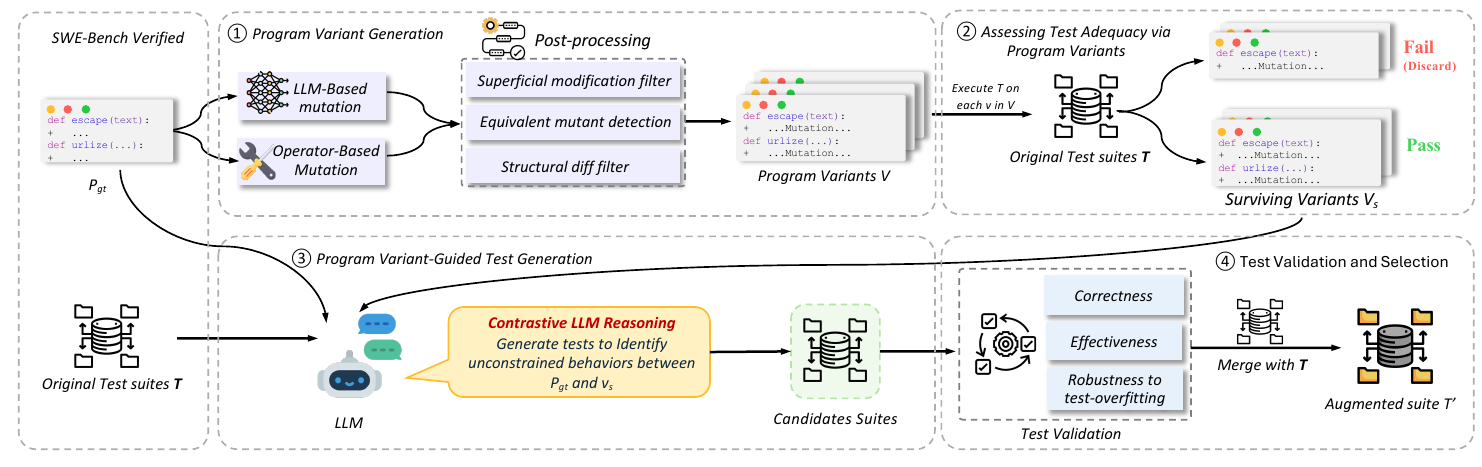}
  \vspace{-1.5em}
  \caption{Overview of the \tool framework.  }
  \label{fig:probe_overview}
  \vspace{-1.em} 
\end{figure*}

To address under-constrained regression test suites in real-world issue 
benchmarks, we propose \tool: a fully automated diagnostic-driven test augmentation 
framework. Given a verified reference patch $P_{gt}$ and its associated 
regression test suite $T$, \tool produces an augmented suite 
$T' \supseteq T$ that more tightly constrains the intended behavior of 
the fix. 
A strong test suite should not only accept the correct fix, it should also \emph{reject} plausible yet incorrect alternatives. \tool operationalizes this intuition by treating semantically modified variants of $P_{gt}$ as \emph{behavioral probes}: if a variant passes all tests in $T$ despite deviating from $P_{gt}$, it exposes a behavioral gap. \tool then generates targeted tests to close that gap.


Figure~\ref{fig:probe_overview} illustrates the overall workflow of 
\tool. Given $P_{gt}$ and $T$, \tool  proceeds through four stages: 
(1)~\textbf{Program Variant Generation} produces semantically 
modified variants of $P_{gt}$ that are syntactically valid and 
interface-compatible (\S\ref{approach:perturbation}); 
(2)~\textbf{Assessing Test Adequacy via Program Variants} executes $T$ against each 
variant, collecting those that pass as evidence of behavioral gaps 
(\S\ref{approach:probing}); 
(3)~\textbf{Program Variant-Guided
Test Generation} generates new tests guided 
by the surviving variants as contrastive signals 
(\S\ref{approach:generation}); and 
(4)~\textbf{Test Validation and Selection} ensures generated tests pass on $P_{gt}$ 
while rejecting at least one surviving variant, admitting only 
behaviorally meaningful additions to $T' = T \cup T_{new}$ 
(\S\ref{approach:validation}).

\begin{table}[htbp]
\centering
\small
\caption{Mutation operators derived from established testing tools and research. }

\label{tab:operators}
\vspace{-0.8em}
\resizebox{\linewidth}{!}{
\begin{tabular}{lllc}
\toprule
Category & Operator & Transformation Example & Source \\
\midrule
\multirow{7}{*}{\shortstack[l]{Predicate \&\\Boolean Logic}}
 & condfalse & condition $\rightarrow$ False & \cite{mutmut} \\
 & condtrue & condition $\rightarrow$ True & \cite{mutmut} \\
 & condflip & negate predicate & \cite{cosmicray} \\
 & boolswap & and $\leftrightarrow$ or & \cite{mutmut,cosmicray,mutpy} \\
 & boollit & replace boolean literal & \cite{mutmut,cosmicray,mutpy} \\
 & eqflip & == $\leftrightarrow$ != & \cite{mutmut,cosmicray,mutpy} \\
 & cmpbound & > $\leftrightarrow$ >= & \cite{mutmut,cosmicray,mutpy} \\
\midrule
\multirow{6}{*}{\shortstack[l]{Arithmetic \&\\Numeric}}
 & numlit & modify numeric literal & \cite{mutmut,cosmicray,mutpy} \\
 & strlit & modify string literal & \cite{mutmut} \\
 & arithop & change arithmetic operator & \cite{cosmicray,mutpy} \\
 & none2zero & None $\rightarrow$ 0 & \cite{mutpy} \\
 & len2zero & len(x) $\rightarrow$ 0 & \cite{mutpy} \\
 & len2one & len(x) $\rightarrow$ 1 & \cite{mutpy} \\
\midrule
\multirow{2}{*}{\shortstack[l]{Return \&\\Default}}
 & retNone & return x $\rightarrow$ return None & \cite{cosmicray,mutpy} \\
 & pass2none & pass $\rightarrow$ return None & \cite{mutpy} \\
\midrule
\multirow{5}{*}{\shortstack[l]{Loop \&\\Iteration}}
 & reverseloop & reverse iteration order & \cite{cosmicray} \\
 & brkcont & break $\leftrightarrow$ continue & \cite{mutmut,cosmicray,mutpy} \\
 & oneloop & limit loop to single iteration & \cite{cosmicray} \\
 & zeroloop & skip loop body & \cite{cosmicray} \\
 & rangepp & modify range bounds & \cite{mutpy} \\
\midrule
\multirow{5}{*}{\shortstack[l]{Data Access \&\\Slicing}}
 & listidx & modify list indexing & \cite{mutpy} \\
 & dictget & dict[k] $\rightarrow$ dict.get(k) & \cite{mutpy} \\
 & slicedel & remove slice operation & \cite{mutpy} \\
 & sliceleft & modify slice start & \cite{cosmicray,mutpy} \\
 & sliceright & modify slice end & \cite{cosmicray,mutpy} \\
\midrule
\multirow{2}{*}{\shortstack[l]{Exception\\Handling}}
 & exctype & change exception type & \cite{cosmicray} \\
 & excswallow & raise E $\rightarrow$ pass & \cite{cosmicray} \\
\midrule
\multirow{5}{*}{\shortstack[l]{Structural\\Transformations}}
 & decdel & delete decorator & \cite{cosmicray,mutpy} \\
 & compfilterdel & [x for x in L if p] $\rightarrow$ [x for x in L] & \cite{cosmicray,mutpy} \\
 & unaryop & modify unary operator & \cite{cosmicray,mutpy} \\
 & bitwiseop & modify bitwise operator & \cite{cosmicray,mutpy} \\
 & augassign & modify augmented assignment & \cite{mutmut,cosmicray,mutpy} \\
\bottomrule
\end{tabular}
}
\vspace{-1.5em}
\end{table}

\subsection{Program Variant Generation}\label{approach:perturbation}

\tool generates a set of \emph{program variants} by applying \textit{controlled mutations} to the reference patch $P_{gt}$. We refer to these as \emph{controlled mutations} because they are confined to the patch region, isolating the behavioral effects of the repair.
These variants act as behavioral probes: variants that pass all tests in $T$ but differ from $P_{gt}$ reveal gaps in the test suite. To focus on behavior introduced by the fix, \tool restricts all transformations to the \emph{patch region}, defined by the location of the modified lines in $P_{gt}$. 
If the modified lines fall inside a function body, the patch region is defined as the enclosing function, allowing access to relevant surrounding logic (e.g., branches and return paths). If the patch spans multiple functions, each function is treated as a separate patch region. For patches outside any function (e.g., module-level statements), the patch region consists of the modified lines.

\begin{figure}[t]

  \centering
  \includegraphics[width=0.9\linewidth]{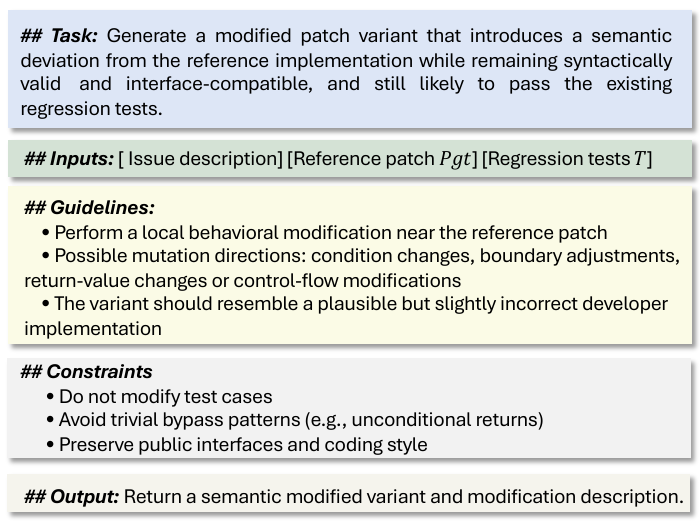}
  \vspace{-1.5em}
  \caption{Simplified Prompt template used in LLM-based Mutation}
  
  \label{fig:llm_mutation}
  \vspace{-1em}
\end{figure}

\tool generates program variants using two complementary strategies and filters them to retain only valid and informative ones, yielding the final set $V$.

\textbf{(i) Operator-based Mutation.} \tool applies 32 predefined mutation operators to the patch region to produce fine-grained, interpretable variations (Table~\ref{tab:operators}), spanning seven categories.  These operators are adapted from widely used Python mutation testing tools, Mutmut~\cite{mutmut} Mutpy~\cite{mutpy} and Cosmic Ray~\cite{cosmicray}. Prior empirical studies have shown that  these operators can effectively induce meaningful semantic changes in Python programs~\cite{guerino2024static}. 

For each patch and each transformation type, \tool performs 10 independent mutation attempts, following preliminary trials where surviving-variant discovery largely saturated within this budget. In each attempt, the transformation is applied only if the patch region contains at least one compatible code location (e.g., \texttt{eqflip} requires a comparison expression). 
When multiple compatible locations are available, \tool randomly selects one and applies the transformation, producing one variant per attempt. 
Each run produces one variant with a single-site modification. 

\textbf{(ii) LLM-based Mutation.} To explore higher-level behavioral in addition to operator-based mutation, \tool prompts an LLM with the issue description, $P_{gt}$, and the existing tests $T$. As shown in Figure~\ref{fig:llm_mutation}, the prompt instructs the model to generate interface-compatible program variants that differ in behavior from $P_{gt}$, such as altering conditional logic, adjusting boundary handling, or restructuring control flow. The existing tests $T$ are provided so that the model can reason about which behaviors are already constrained and target deviations more likely to survive the current suite. The prompt further excludes trivial bypass patterns (e.g., unconditional returns) and pure refactoring, as such changes do not represent meaningful behavioral deviations. Unlike operator-based mutation, LLM-based mutation can introduce coordinated changes across multiple program elements, such as jointly modifying a predicate and its corresponding return value. For each instance, we query the LLM 10 times with the same prompt. If a response duplicates a previously generated variant, it is discarded and re-queried to ensure diversity. 

\begin{algorithm}[t]
\caption{Identifying Surviving Variants}
\label{alg:prober}
\small
\begin{algorithmic}[1]
\Require Program variant set  $V$, regression test suite $T$
\Ensure Surviving variant set $V_s$

\State $V_s \gets \emptyset$

\ForAll{$v \in V$}
    \State $\mathit{result} \gets \textsc{RunTests}(T, v)$
    \If{$\mathit{result}$ contains environment errors or timeouts}
        \State \textbf{continue}
    \EndIf
    \If{$\mathit{result}$ passes all tests in $T$}
        \State $V_s \gets V_s \cup \{v\}$
    \EndIf
\EndFor

\State \Return $V_s$
\end{algorithmic}
\end{algorithm}

\paragraph{Post-processing.} Although our mutation operators are designed to induce genuine semantic changes, and our LLM-based guidelines explicitly encourage behaviorally divergent program variants, it is still difficult to guarantee that all generated variants represent meaningful behavioral deviations. Some may be syntactically different but semantically equivalent to the oracle patch, which would create false signals of test weakness and introduce noise into downstream test generation. To ensure that only genuinely divergent variants guide test synthesis, \tool applies three layers of filtering. First, duplicates and variants that modify only superficial elements,such as comments, formatting, docstrings, and logging messages, are discarded by comparing normalized AST with Python's \texttt{ast} library. Second, following Tian et al.~\cite{tian2024large}, who demonstrated that LLMs can effectively distinguish equivalent from non-equivalent mutants, \tool use their prompting template to screen each variant and remove those classified as equivalent rewrites. Third, a structural diff filter, implemented as a normalized-AST comparison, retains only variants whose modifications align with the oracle patch in scope (same modified files, no additional hunks) and whose normalized code changes are not reducible to identifier renaming, message reformatting, or code motion. After these steps, the remaining variants form $V$.


\subsection{Assessing Test Adequacy via Program Variants}\label{approach:probing}

\tool assesses the adequacy of $T$ by executing each program variant $v \in V$ against the full regression test suite and identifying those that pass. A variant that passes all tests in $T$ but violates the intended behavior of $P_{gt}$ is a \emph{surviving variant}, indicating that $T$ could not distinguish it from the reference patch.

Algorithm~\ref{alg:prober} describes this procedure. For each variant $v \in V$, \tool executes $T$ and records the result (Lines~2--3). Executions that fail due to environment errors or timeouts are discarded (Lines~4--6), as they are not attributable to program logic. Variants that fail at least one test are excluded,
since their behavioral deviation is already detected by $T$. Finally, variants that pass all tests are collected into the \emph{surviving variant set} $V_s$ (Lines~7--9).

$V_s$ highlights the potential behavioral gaps in $T$, which are the program behavior that the existing tests fail to constrain. \tool uses these surviving variants as diagnostic signals to guide targeted test generation in the next stage.


\begin{figure}[t]

  \centering
  \includegraphics[width=0.9\linewidth]{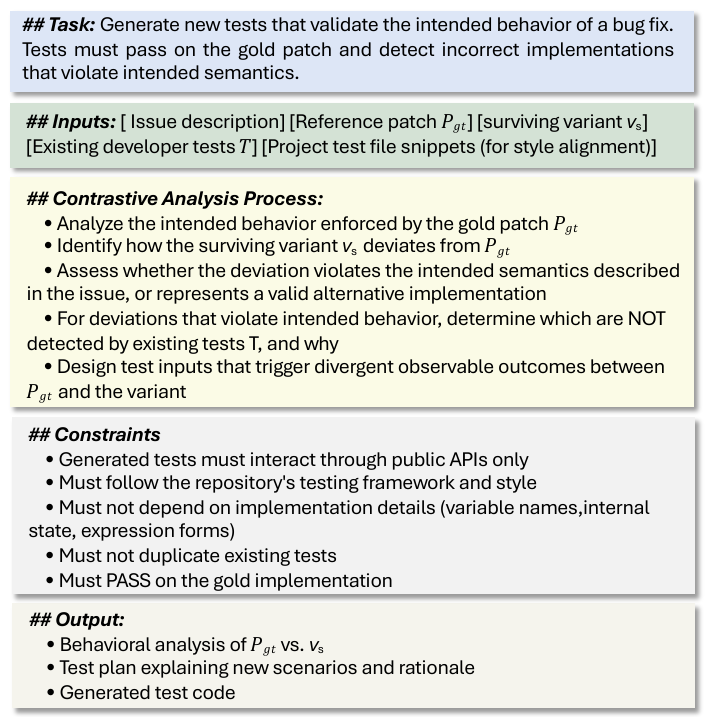}
  \vspace{-1em}
  \caption{Simplified prompt template used in targeted test generation.
  } 
  \label{fig:testgeneration}
\end{figure}

\subsection{Program Variant-Guided Test Generation}\label{approach:generation}

\tool generate additional test cases to address the behavioral 
gaps revealed by $V_s$. \tool 
uses surviving program variants as guidance: each variant identifies a concrete behavioral difference that $T$ fails to detect, providing a target for test generation.

\tool employs an LLM-based test generator that takes as input
$P_{gt}$, a surviving variant $v_s$, the existing
tests $T$, and project test file snippets for style alignment.
We adopt a \emph{contrastive reasoning} strategy~\cite{chia2023contrastive}, implemented through a structured prompt (Figure~\ref{fig:testgeneration}).
The test generator first analyzes how $v_s$ deviates from the intended behavior of $P_{gt}$, then identifies scenarios where existing tests fail to distinguish them, and finally produces targeted regression tests for those scenarios.

To reduce the risk that generated tests depend on implementation artifacts rather than behavioral differences, \tool restricts all generated tests to interact only through the program’s public methods. This design choice encourages generated tests to validate externally observable behavior instead of superficial implementation details. By targeting such inputs or execution conditions, the generated tests are more likely to reveal behaviors that are not sufficiently constrained by the current test suite.  

To ensure the generated tests are meaningful and consistent with the intended behavior, \tool enforces the following two principles:

\noindent\textit{(1) Behavioral Differentiation.} Each generated test must produce different observable outcomes when executed on $P_{gt}$ and at least one surviving variant in $V_s$, thereby exposing semantic discrepancies.

\noindent\textit{(2) Intent Alignment.} $P_{gt}$ serves as the authoritative oracle: a generated test is accepted only if it passes on $P_{gt}$, ensuring consistency with the intended behavior of the fix. 

Because a single $v_s$ may expose multiple behavioral differences from $P_{gt}$, \tool allows the LLM to produce multiple test cases per contrastive query, each targeting a different aspect of the deviation. Finally, \tool produces a candidate test set 
$T_{candidate}$ that targets behavioral regions previously overlooked by $T$.

\begin{table}[htbp]
\centering
\vspace{-1em}
\caption{Behavior-preserving transformations used in robustness validation~\cite{hort2025semantic}.}
\label{tab:semanticpreserving}
\vspace{-1em}
\small
\resizebox{\columnwidth}{!}{
\begin{tabularx}{\columnwidth}{p{2.7cm} X}
\toprule
\textbf{Transform}   & \textbf{Description} \\
\midrule
Replace names & Consistently renames functions, classes, variables, parameters, etc.\ while preserving scope. \\
Swap operands & Swaps the operands of simple binary comparisons, e.g.\ \texttt{a < b} $\to$ \texttt{b > a}, \texttt{x == y} $\to$ \texttt{y == x}. \\
Reorder statements (*)& Reorders independent local statements that have no data dependencies. \\
Split if-conditions & Splits compound conditions into nested if-statements while preserving short-circuit semantics. \\
Combine if-statements & Merges consecutive if-statements into a single compound condition, preserving evaluation order. \\
ConExpr $\leftrightarrow$ If-else & Converts ternary conditional expressions into multi-line if-else blocks and vice versa. \\
For $\leftrightarrow$ While (*)& Transforms for-loops into equivalent while-loops and vice versa. \\
Arithmetic transform & Converts augmented assignments to expanded form and vice versa, e.g.\ \texttt{x += 1} $\leftrightarrow$ \texttt{x = x + 1}. \\
ListComp $\leftrightarrow$ ForLoop (*)& Rewrites list comprehensions into equivalent explicit for-loops and vice versa. \\
Boolean simplify (*)& Simplifies boolean expressions, e.g.\ \texttt{x is True} $\to$ \texttt{x}, \texttt{not (x > y)} $\to$ \texttt{x <= y}. \\
FString $\leftrightarrow$ Format & Converts f-strings into equivalent \texttt{.format()} calls and vice versa. \\
\bottomrule
\end{tabularx}
}
\vspace{-1em}
\end{table}
\vspace{-1em}

\subsection{Test Validation and Selection}\label{approach:validation}

Before incorporating any candidate test into the augmented suite, \tool validates it against three criteria. This validation is designed to ensure that retained tests are consistent with the intended behavior of $P_{gt}$, add genuinely new behavioral constraints, and do not overfit superficial implementation details. Each test must therefore satisfy three criteria:

\noindent\textbf{Correctness.} A generated test must pass on $P_{gt}$, 
ensuring consistency with the intended behavior of the fix. Tests 
that fail on $P_{gt}$ are discarded.

\noindent\textbf{Effectiveness.} \tool retains a test only if it exposes new behavioral discrepancies. It executes each candidate test on the surviving variants $V_s$ and keeps the test only if it fails on at least one variant, indicating that it constrains behavior not covered by $T$.

\noindent\textbf{Robustness to Test Overfitting.}
Prior studies have shown that LLM-generated code can be sensitive to superficial features of the input context, such as identifier names and code structure~\cite{gao2023two,wang2023does}. In the test generation setting, this means generated tests may inadvertently encode implementation-specific properties, such as specific variable names or control-flow patterns, rather than observable program behavior. As a result, they may incorrectly reject semantically equivalent implementations. To detect such overfitting tests, \tool applies behavior-preserving transformations to $P_{gt}$, including identifier renaming, operand swapping, and control-flow restructuring (Table~\ref{tab:semanticpreserving}). Because these transformations preserve semantics, a valid test should pass on all transformed variants. Tests that fail on any such variant are discarded, as they likely depend on superficial artifacts rather than program behavior.  In addition to transformation-based filtering, PROBE uses an LLM-based screening step that classifies each retained test as either a valid behavioral strengthening or an implementation-specific check tied to the oracle patch. Tests classified as implementation-specific are discarded.

\section{Evaluation}\label{sec:empirical}

In this section, we evaluate PROBE by answering four research questions (RQs).

\phead{Benchmark.}
We conduct our evaluation on SWE-bench
Verified~\cite{swebench_verified}, a curated benchmark of 500
real-world software issues collected from widely used open-source Python repositories. We select this benchmark for two reasons. First, its evaluation relies entirely on the provided regression tests $T$. The strength of $T$ directly determines whether benchmark results faithfully reflect patch correctness, making test adequacy a first-class concern. Second, its official evaluation harness executes tests in isolated Docker environments, enabling reproducible experimentation at scale.

\phead{Implementation.} 
In \tool, the LLM is used only to propose candidate variants and tests, while post-processing and execution-based validation determine which artifacts are retained. This separation decouples \tool from any specific LLM: any code-capable LLM that can follow the prompt-based input-output format can instantiate the candidate-generation component. Model choice may affect the diversity and quality of generated candidates, but the subsequent filtering, test execution, and robustness validation are independent of the underlying model. In our evaluation, we use GPT-5-mini (\texttt{gpt-5-mini-2025-08-07}, reasoning level: medium) through the OpenAI API because it provides a practical balance between code reasoning ability and cost at the scale of SWE-bench Verified.
For operator-based mutation, all 32 operators are applied to compatible sites within the patch region, with each operator executed for 10~attempts using randomly sampled transformation sites. For LLM-based mutation, 10 independent variants are generated per instance; duplicate responses are discarded and re-queried. 

\vspace{-0.5em}
\subsection*{RQ1: What behavioral gaps exist in SWE-bench regression tests?}


\phead{Motivation.} Regression test suites may fail to fully constrain the intended behavior of a fix: a program variant can pass all tests while still violating the intended semantics of the reference patch. Such cases indicate that the benchmark silently accepts incorrect implementations. In this RQ, we study how often this occurs and which behavioral gaps enable such variants to survive, which is critical for assessing the reliability of existing benchmarks. 



\phead{Approach.} We apply both operator-based and LLM-based mutation (\S\ref{approach:perturbation}) to all 500 instances in SWE-bench Verified. For each instance, we execute the existing regression tests $T$ against each generated variant and collect the surviving variant set $V_s$. 

First, to measure \textit{\textbf{how often test suites fail to fully constrain program behavior}} in SWE-bench Verified, we count the number of program variants that survive the existing tests. Specifically, we report the number of surviving variants and the number of affected issues for each variant generation method, including operator-based mutation, LLM-based mutation, and their combination after deduplication. 

Second, to understand \textit{\textbf{why program variants survive}}, we analyze the types of behavioral gaps that allow them to pass existing tests. We first examine the distribution of surviving operator-based variants across mutation categories (Table~\ref{tab:semantic-survivors}) to identify which mutations are most likely to evade detection.   For LLM-based surviving variants, whose changes are not constrained to predefined operators, we instead classify each surviving variant by its top-level modified AST node type following the previous method~\cite{pan2009toward} (Table~\ref{tab:llm_change_patterns}), to characterize the structural nature of changes that evade existing tests. We then conduct a qualitative analysis on a random 20\% sample of instances with surviving variants, following prior mutation-testing studies~\cite{straubinger2024empirical}. 
The first author applies open coding to the representative surviving variants, the reference patch, and the corresponding tests to derive an initial set of categories with names and definitions. The second author then independently assigns each instance to these categories. We report Cohen's
over the two independent label sets to quantify the reliability of this manual categorization, together with the number of disagreements, which are resolved through discussion, after which the category names and definitions are finalized.

\begin{table}[t]
\centering
\small
\caption{Surviving program variants and affected instances by the two mutation
strategies on SWE-bench Verified (500 instances total).}
\vspace{-1em}
\label{tab:test_weakness}
\begin{tabular}{lcc}
\toprule
 & Operator-based & LLM-based \\
\midrule
Affected SWE-bench instances & 50 / 500 (10.0\%) & 380 / 500 (76.0\%) \\
Surviving program variants & 209 & 1{,}915 \\
Avg. variants per instance & 4.2 & 5.0 \\
\midrule
Combined unique instances & \multicolumn{2}{c}{385 / 500 (77.0\%)} \\
\bottomrule
\end{tabular}
\vspace{-1.5em}
\end{table}




\noindent\textbf{Results and Discussions.}
\textbf{\textit{77\% of SWE-bench Verified instances have at least one surviving program variant that passes all existing regression tests.}}
Table~\ref{tab:test_weakness} reports the number of surviving program variants and affected instances for each mutation strategy. 385 of 500 instances (77.0\%) contain at least one surviving program variant, showing that under-constrained tests are widespread in the benchmark rather than limited to a few edge cases. LLM-based mutation accounts for the majority, affecting 380 instances with 1,915 variants. Meanwhile, operator-based mutation affects 50 instances with 209 variants, of which 5 are not covered by LLM-based variants. LLM-based mutation also produces more variants per affected instance (5.0 vs 4.2), suggesting that it explores a broader space of behavioral deviation per issue.

\begin{table}[t]
\centering
\small
\caption{Surviving variants not killed by developer-written tests, grouped by semantic category with their contributing operators.}
\label{tab:semantic-survivors}
\vspace{-1em}
\begin{tabular}{l l c}
\toprule
Semantic Category & Operator & {\# Surviving} \\
\midrule

\multirow{7}{*}{Predicate \& Boolean Logic (52.2\%)}
& condfalse & 55 \\
& condtrue  & 16 \\
& eqflip    & 13 \\
& boollit   & 9 \\
& cmpbound  & 7 \\
& condflip  & 5 \\
& boolswap  & 4 \\

\hdashline

\multirow{6}{*}{Arithmetic \& Numeric (32.1\%)}
& len2zero  & 17 \\
& numlit    & 14 \\
& strlit    & 14 \\
& len2one   & 13 \\
& arithop   & 6 \\
& none2zero & 3 \\

\hdashline

\multirow{2}{*}{Loop \& Iteration (6.7\%)}
& reverseloop & 11 \\
& brkcont     & 3 \\

\hdashline

Exception Handling (4.3\%)  & exctype & 9 \\

Return \& Default (3.3\%)  & retNone & 7 \\
 
Structural Transfromations (1.4\%)        & decdel  & 3 \\

\midrule
\textbf{Total} &  & \textbf{209} \\
\bottomrule
\end{tabular}
\vspace{-0.8em}
\end{table}

\noindent\textit{\textbf{Conditional logic is the dominant source of surviving program variants in both strategies (52.2\% operator-based, 54.3\% LLM-based).}} 
Table~\ref{tab:semantic-survivors} breaks down the 209 surviving operator-based variants by mutation category. \textit{Predicate and Boolean Logic} mutations are the most common types of operators, accounting for 109 of 209 variants (52.2\%), with \texttt{condfalse} alone contributing 55. The finding indicates that many conditional branches in patch regions are not exercised by any test input. \textit{Arithmetic and Numeric} mutations follow at 67 variants (32.1\%). Together, these two categories account for over 84\% of all surviving variants. The remaining categories contribute 33 variants combined (15.7\%).

\begin{table}[t]
\centering
\small
\caption{Change patterns in LLM-based surviving variants, classified by the top-level modified AST node following prior taxonomy of bug fix patterns~\cite{pan2009toward}.}
\label{tab:llm_change_patterns}
\vspace{-0.8em}
\resizebox{\columnwidth}{!}{%
\begin{tabular}{l c r}
\toprule
\textbf{Change Pattern} & \textbf{AST Node Types} & \textbf{\# Surviving} \\
\midrule
Condition Modification (54.3\%) & If, Compare & 1,040 \\
Function Modification (16.9\%) & FunctionDef, ClassDef & 324 \\
Assignment Modification (11.9\%) & Assign, AugAssign & 228 \\
Return Modification (6.9\%) & Return & 132 \\
Expression/Call Modification (4.0\%) & Call, Expr & 77 \\
Loop Modification (2.8\%) & For, While & 53 \\
Exception Handling Modification (1.7\%) & Try, Raise & 33 \\
Other (1.5\%) & -- & 28 \\
\midrule
\textbf{Total} & & 1,915 \\
\bottomrule
\end{tabular}
}
\vspace{-1em}
\end{table}

Table~\ref{tab:llm_change_patterns} shows LLM-based surviving variants by classifying each change according to the top-level modified AST node type, following prior taxonomy of bug fix patterns~\cite{pan2009toward}. \textit{{Condition Modification}} is the most frequent pattern (54.3\%), consistent with the dominance of conditional mutations observed in operator-based variants. The results indicate that test suites often fail to constrain branch behavior. 
Moreover, LLM-based mutations introduce more complex changes, including \textit{{Function Modification}} (16.9\%) and \textit{{Assignment Modification}} (11.9\%), which typically span several statements or broader code regions. These patterns are difficult to capture with first-order mutants, suggesting that LLM-based mutations can expose higher-level behavioral gaps beyond localized changes.

Based on the manual analysis by two authors (Cohen's $k=\textbf{0.94}$), we identify four recurring patterns in \textbf{\textit{how existing tests fail to constrain surviving variants}}. Insufficient Input Space Exploration 
is the most common (55.8\%), and Missing Environmental Context is the least (7.8\%). We discuss each pattern in detail below: 

\noindent\textit{\textbf{Pattern 1:  Insufficient Input Space Exploration (43, 55.8\%).}}
Tests check only a small set of inputs and miss other valid cases, such as boundary values or alternative configurations.
For example, in \texttt{django-11239}, the fix ensures that each SSL parameter is passed correctly to the subprocess. However, the test checks only one configuration, \texttt{sslmode=verify-ca}, and does not cover other SSL settings.
This is the most common pattern, showing that many tests verify only a few representative inputs. 

\noindent\textit{\textbf{Pattern 2: Partial Patch Path Coverage  (15, 19.5\%).}}
Tests cover the primary execution path but skip alternative paths that the patch also modifies, such as fallback logic or default implementations.
For example, in \texttt{django-11095}, the fix supports both a custom \texttt{get\_inlines()} implementation and the default behavior when no custom implementation is provided. The test checks only the custom case and does not cover the default one.
This pattern is common when a fix changes both the default behavior and an overridden behavior, but the test exercises only one path.

\noindent\textit{\textbf{Pattern 3: Weak Assertions (13, 16.9\%).}}
Tests verify only the visible output and miss deeper properties of the expected behavior, such as type, metadata, or structure.
For example, in \texttt{django-9296}, the fix ensures that iterating over \texttt{Paginator} yields \texttt{Page} objects rather than raw lists. However, the test only converts the result to a list and checks its contents, so it cannot detect whether the returned values are actually \texttt{Page} objects. This results in three incorrect variants that return raw data still pass.
This pattern allows a patch to satisfy the test at the output level while violating the interface that downstream code relies on.

\noindent\textbf{\textit{Pattern 4: Missing Environmental Context (6, 7.8\%).}}
Tests execute in a clean setup and fail to capture behaviors that depend on existing environment context.
For example, in \texttt{django-10973}, the test uses a mocked subprocess with no pre-existing \texttt{PGPASSWORD} and always assumes success. A surviving variant fails when \texttt{PGPASSWORD} is already present and also skips error checking, but the test fails to detect issues in both variants.
Although this is the least common pattern, it is particularly risky in practice because the patch may pass in CI yet fail silently in production. 

\begin{findings}
Under-constrained tests are widespread: 77\% of instances admit
at least one incorrect variant that still passes all tests. Most
gaps come from insufficient input space exploration (55.8\%) and
missing execution paths (19.5\%). Overall, tests tend to validate
the immediate fix rather than the full behavior it affects.
\end{findings}
\nopagebreak[4]

\subsection*{\textbf{RQ2: How do the augmented tests improve the coverage of the regression test
suite?} }

\textbf{Motivation.} RQ1 shows that many SWE-bench Verified instances remain under-constrained: program variants can still pass the original regression suite. We therefore investigate whether the augmented tests generated by \tool improve coverage of the fix-relevant behavior that the original suite fails to exercise.

\phead{Approach.}
We compare the original test suite $T$ with the augmented suite $T' = T \cup T_{new}$, where $T_{new}$ contains validated generated tests. Our evaluation focuses on whether $T'$ better constrains the fix-relevant behavior that allowed variants to survive under $T$. 
We measure this along two complementary dimensions: (1) structural coverage and (2) assertion strength. For structural coverage, we report line coverage and branch coverage over the patch region, computed as the percentage of executed lines and branches within the modified code. These metrics quantify whether the augmented suite exercises additional execution paths introduced or affected by the fix~\cite{namin2009influence}.

\begin{table}[htbp]
\centering
\footnotesize
\setlength{\tabcolsep}{4pt}
\renewcommand{\arraystretch}{1.03}
\caption{Per-issue static characteristics of developer-written tests versus generated tests.}
\vspace{-0.8em}
\label{tab:assetion}
\begin{tabular}{lccc}
\toprule
\textbf{Metric} & \textbf{Original} & \textbf{Generated} & \textbf{Delta} \\
\midrule
Assertion Number & 2.31 & 5.18 & +124.2\% \\
Assertion Density & 0.22 & 0.30 & +36.4\% \\
Assertion Types & 1.56 & 3.14 & +101.2\% \\
\bottomrule
\end{tabular}
\vspace{-1.5em}
\end{table}

\begin{figure}[htbp]
  \centering
\includegraphics[width=0.62\linewidth]{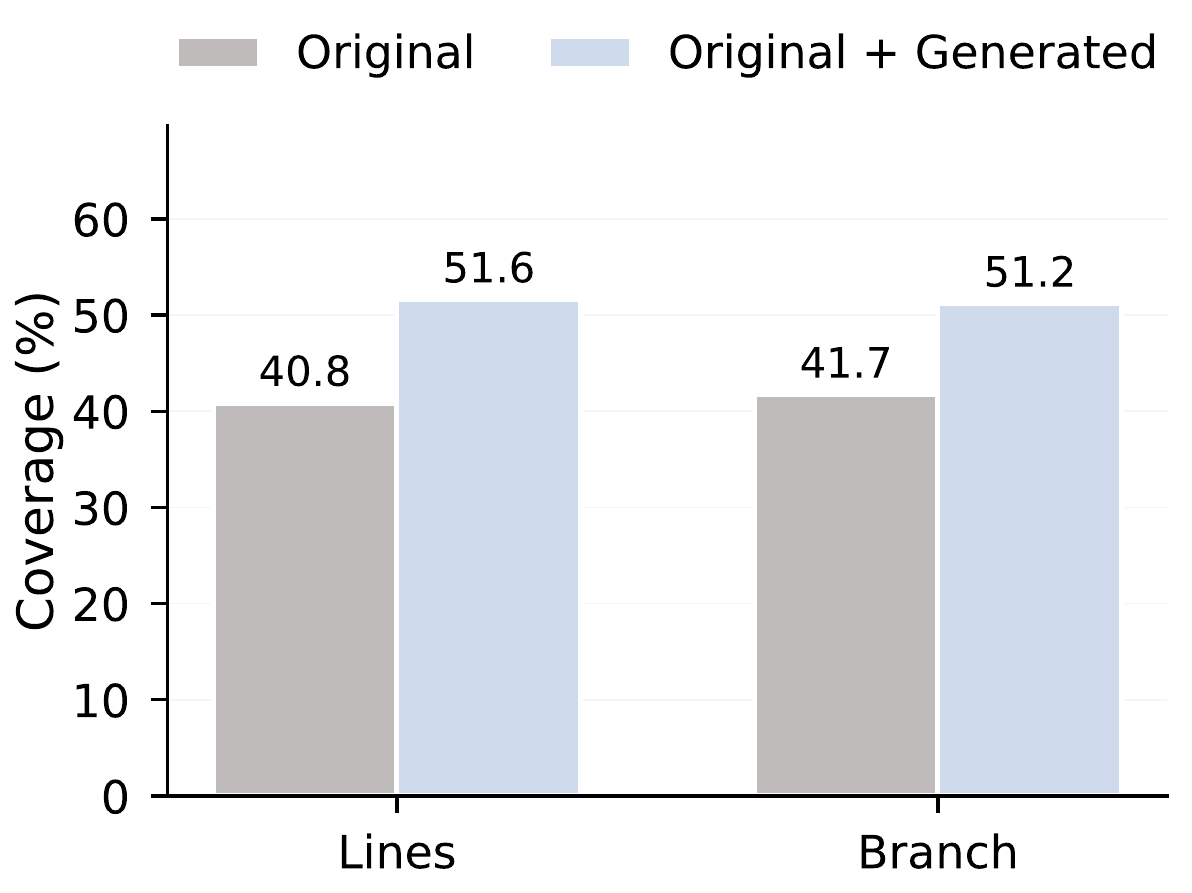}
\vspace{-1em}
  \caption{ Coverage comparison between original and augmented test suites.}
  \vspace{-1em}
  \label{fig:coverage}
  
\end{figure}

For assertion strength, we report three metrics, all computed automatically by parsing each test: (i) assertions number, defined as the average number of assertion statements in each test suite; (ii) assertion density, defined as the ratio of assertion statements to total statements within a test suite, capturing how much of the test is devoted to behavioral checking rather than setup or execution; and (iii) assertion type diversity, defined as the number of distinct assertion types used, where a type denotes the kind of property checked, such as equality (assertEqual or ==), exception (assertRaises), or membership (assertIn), reflecting the variety of behavioral properties being validated. Together, these metrics characterize how thoroughly the tests check the executed behavior, beyond simply increasing coverage~\cite{zhang2015assertions, catolino2019experience}. 
To statistically confirm that the improvements are consistent across issues, we further apply paired Wilcoxon signed-rank tests and report rank-biserial correlation as the effect size. This non-parametric test is appropriate because our per-issue measurements are paired and not normally distributed~\cite{statisticaltests}

\phead{Results and Discussions.} 
\tool generates 1,316 candidate test cases across 236 test-weak instances. After  removing tests that depend on implementation-specific behavior (\textit{Robustness} validation in Section ~\ref{approach:validation}), 1,014 tests across 211 instances are retained for augmentation. For each issue, we compare these retained tests with the developer-written test suite, including both newly added tests and modifications to existing ones. 
The augmented test suites substantially expand coverage over the patch region (i.e., fix-relevant code). As shown in Figure~\ref{fig:coverage}, across the 211 instances, average line coverage increases from 40.8\% to 51.6\%, and branch coverage from 41.7\% to 51.2\%. The improvements are statistically significant under paired Wilcoxon signed-rank tests ($p < 0.001$ and $p\approx0.02$, respectively), and all non-zero paired differences favor augmentation. This indicates that the generated tests exercise previously untested execution paths.

More importantly, the improvement is not limited to executing more code.  The generated tests also strengthen behavioral constraints through 
more, denser, and more diverse assertions, each associated with stronger fault-detection capability in prior work~\cite{zhang2015assertions, catolino2019experience, togll}. 
To characterize the checking strength contributed by the generated tests themselves, we further compare their assertion structure with that of the developer-written tests. As shown in Table~\ref{tab:assetion}, they contain more than twice as many assertions per test suite (5.18 vs.\ 2.31), higher assertion density (0.30 vs.\ 0.22), and greater assertion type diversity (3.14 vs.\ 1.56 unique types). These differences are statistically significant under paired Wilcoxon signed-rank tests (all p < 0.001), with medium-to-large effect sizes  $r_{rb}=0.55, 0.48$ and $0.52$, respectively. Together, these results indicate that the additional coverage is accompanied by richer behavioral checks, rather than superficial execution alone.

\begin{findings}
\tool improves coverage of under-constrained regression suites by targeting fix-relevant code paths missed by developer-written tests. The augmented suite increases patch-region line coverage by 10.8\% and branch coverage by 9.5\%, while also providing richer assertions over the covered code paths.
\end{findings}

\subsection*{\textbf{RQ3: How does stronger testing change repair agent evaluation?}}




\phead{Motivation.} RQ1 and RQ2 show that \tool exposes test gaps and generates targeted tests that improve coverage. In this RQ, we examine how these stronger tests affect the evaluation of repair agents. Specifically, we re-run patches produced by these agents on the augmented test suite to determine whether incorrect patches that previously passed are now detected. 


\begin{table*}[t]
\centering
\caption{Impact of \textcolor{orange!80!black}{PROBE} augmented tests on repair-agent evaluation, with the UTBoost baseline. Resolved\% gives the reported rate (\%) before augmentation (Orig.) and after augmenting with UTBoost  and \textcolor{orange!80!black}{PROBE}  ($\Delta$ = change from Orig.). Killed(\#) counts patches that passed $T$ but fail a test added by each method; \emph{Overlap} denotes patches killed by both UTBoost and \textcolor{orange!80!black}{PROBE}. Adjusted Rank is the rank under \textcolor{orange!80!black}{PROBE}-augmented tests evaluation.}
\label{tab:tool-patch-validation}
\vspace{-0.8em}
\small
\setlength{\tabcolsep}{4pt}
\resizebox{\textwidth}{!}{%
\begin{tabular}{lccccccccc}
\toprule
\multirow{2}{*}{\textbf{Repair Agent}} & \multicolumn{3}{c}{\textbf{Resolved (\%)}} & \multicolumn{3}{c}{\textbf{Killed (\#)}} & \multirow{2}{*}{$T'_{op}$} & \multirow{2}{*}{$T'_{llm}$} & \multirow{2}{*}{\textbf{Adjusted Rank}} \\
\cmidrule(lr){2-4} \cmidrule(lr){5-7}
 & Orig. & +UTBoost ($\Delta$) & \textbf{\textcolor{orange!80!black}{+PROBE} ($\Delta$)} & UTBoost & Overlap & \textbf{\textcolor{orange!80!black}{PROBE}} & & & \\
\midrule
live-SWE-agent (Claude 4.5 Opus medium)~\cite{xia2025live} & 79.2 & 77.0 (-2.2) & \textbf{75.0 (-4.2)} & 11 & 0 & \textbf{21} & 3 & 20 & 1st \\
Sonar Foundation Agent (Claude 4.5 Opus)~\cite{Sonar} & 79.2 & 77.8 (-1.4) & \textbf{75.0 (-4.2)} & 11 & 0 & \textbf{21} & 3 & 21 & 1st \\
Trae Doubao Seed Code~\cite{gao2025trae}  & 78.8 & 75.6 (-3.2) & \textbf{72.4 (-6.4)} & 15 & 2 & \textbf{32} & 5 & 31 & 3rd \\
live-SWE-agent (Gemini 3 Pro Preview)~\cite{xia2025live} & 77.4 & 75.4 (-2.0) & \textbf{71.4 (-6.0)} & 11 & 3 & \textbf{30} & 3 & 30 & 4th \\
Atlassian Rovo Dev~\cite{Rovo} & 76.8 & 73.4 (-3.4) & \textbf{68.8 (-8.0)} & 16 & 2 & \textbf{40} & 5 & 39 & 8th\textcolor{red}{$\downarrow$} \\
EPAM AI (Claude 4 Sonnet)~\cite{epam} & 76.8 & 74.6 (-2.2) & \textbf{69.4 (-7.4)} & 12 & 2 & \textbf{37} & 4 & 35 & 5th\textcolor{ForestGreen}{$\uparrow$} \\
ACoder~\cite{acoder} & 76.4 & 74.6 (-1.8) & \textbf{69.0 (-7.4)} & 14 & 1 & \textbf{37} & 5 & 35 & 7th \\
Warp~\cite{Warp} & 75.6 & 72.2 (-3.4) & \textbf{69.4 (-6.2)} & 15 & 0 & \textbf{31} & 4 & 30 & 5th\textcolor{ForestGreen}{$\uparrow$} \\
TRAE~\cite{gao2025trae} & 75.2 & 73.4 (-1.8) & \textbf{66.2 (-9.0)} & 12 & 3 & \textbf{45} & 6 & 43 & 10th\textcolor{red}{$\downarrow$} \\
Harness AI~\cite{harness} & 74.8 & 72.2 (-2.6) & \textbf{67.8 (-7.0)} & 12 & 1 & \textbf{35} & 5 & 33 & 9th\textcolor{ForestGreen}{$\uparrow$} \\
\midrule

\textbf{Avg./Total.} &  & -2.4\,{\scriptsize(Avg.)} & \textbf{-6.6}\,{\scriptsize(Avg.)} & 129 & 14 & \textbf{329} & 38 & 317 &  \\

\bottomrule
\end{tabular}
}
\vspace{-0.8em}
\end{table*}

\begin{table}[!t]
\centering
\captionof{table}{Consistency of \tool across three independent runs of the test-generation and re-evaluation pipeline.}
\label{tab:nondeterminism}
\vspace{-0.8em}
\small
\begin{tabular}{lcccc}
\toprule
Run & Validated tests & Instances & Avg.\ Drop (\%) & Patches killed \\
\midrule
Run 1 & 1,014 & 211 & 6.58 & 329 \\
Run 2 & 1,042 & 214 & 7.06 & 353 \\
Run 3 & 973 & 203 & 6.24 & 312 \\
\bottomrule
\end{tabular}
\vspace{-1em}
\end{table}

\phead{Approach.}
We re-evaluate the top-10 repair agents on the SWE-bench Verified leaderboard~\cite{leaderboard} whose submitted patches are publicly available. For each agent-generated patch, we execute both the original tests $T$ and the augmented suite $T' = T \cup T_{new}$ produced by \tool. A patch is counted as \textit{killed} if it originally passed $T$ but fails at least one test in $T_{new}$. We then compute each agent's resolved rate under $T'$ and compare it against the original rate. To understand the contribution of each mutation strategy, we further report kills separately for tests derived from the operator-based mutation pipeline ($T'_{op}$) and the LLM-based mutation pipeline ($T'_{llm}$).
 We also compare \tool with UTBoost~\cite{yu2025utboost}, the most directly comparable and openly available framework for SWE-bench test augmentation, under the same setup and evaluation.
We execute UTBoost using the same model as \tool (GPT-5-mini) and run end-to-end to generate an augmented test suite. Both are then evaluated on the same set of top-10 agent patches, with per-agent resolved-rate changes, killed patches, and overlap reported to assess complementarity. Finally, to assess reproducibility under LLM nondeterminism, we analyze the consistency of \tool's results across three independent runs of the full test-generation and re-evaluation pipeline.

\phead{Results.} 
Table~\ref{tab:tool-patch-validation} shows that the augmented tests consistently change evaluation outcomes for all ten repair agents. Resolved-rate drops range from 4.2\% to 9.0\%, with 21 to 45 previously accepted patches per agent now rejected. The two mutation strategies are complementary: LLM-based tests account for most of the detected failures, while operator-based tests contribute an additional 3 to 6 unique failures per agent that LLM-based tests do not catch. 
These corrections also change the leaderboard. \textit{Atlassian Rovo Dev}, \textit{ACoder}, and \textit{TRAE} drop in rank, while \textit{EPAM AI}, \textit{Warp}, and \textit{Harness AI} move up. This change indicates that the original test suites accepted a non-trivial number of incorrect patches, inflating the reported performance of some agents. Overall, the results show that stronger tests provide a more reliable comparison between repair agents.

{\textbf{Comparison with baseline.}} As Table~\ref{tab:tool-patch-validation} shows, \tool rejects substantially more previously passing patches than UTBoost across all ten agents: the average resolved-rate drop is 6.6\,pp versus 2.4\,pp ($\sim2.75\times$). In total, \tool kills 329 previously accepted patches versus 129 for UTBoost, with only 14 killed by both. The small overlap indicates that the two methods detect largely different incorrect patches, with \tool exposing substantially more. This complementarity reflects their different detection sources (§\ref{sec:related}): UTBoost observes weaknesses through agent-submitted patches, whereas \tool mutates the oracle patch to expose benchmark gaps independently of any agent.

{\textbf{Reproducibility under LLM nondeterminism.}} Across the three independent runs in Table~\ref{tab:nondeterminism}, \tool retains 973--1{,}042 tests over 203--214 instances and lowers resolved rates by 6.24--7.06\% on average. These results suggest that, although there is some variation, the benchmark weaknesses exposed by \tool are not specific to a single LLM run.

\enlargethispage{\baselineskip}

\begin{findings}
\textbf{Findings:} \tool's augmented tests reduce resolved rates by 4.2\% to 9.0\% across all 10 agents, with some repair agents experiencing larger drops. Compared with UTBoost, \tool rejects more previously accepted patches with limited overlap, showing complementary weakness detection. The same re-evaluation trend appears across three independent LLM runs.
\end{findings}
\vspace{-0.5em}

\vspace{-0.8em}
\subsection*{\textbf{RQ4: What behavioral checks do the augmented tests introduce?}}
\vspace{-0.4em}


\phead{Motivation.}
RQ3 shows that the augmented tests reject many previously accepted patches. To understand what makes these tests effective, 
 we analyze the behavioral checks they add and how each exposes flaws missed by original tests.



\noindent\textbf{Approach.} We randomly sample  66 (20\%) newly rejected patches in RQ3 for manual analysis.
For each sample, we examine the original tests, the oracle patch, and the augmented tests that cause the rejection to understand why the patch fails. By comparing the assertions, inputs, and execution paths introduced by the augmented tests against those in the original suite, we identify the behavioral property that the augmented tests newly constrain. Two authors independently analyze the sampled patches. The first author develops an initial set of categories by open-coding all samples, identifying recurring types of behavioral checks that the augmented tests introduce. The second author then independently assigns each sample to one of the proposed categories. Disagreements are resolved through discussion, after which the category definitions are finalized. We report Cohen's $\kappa$ on the independent assignments to assess inter-rater agreement. Both authors also assess whether each rejection reflects genuine semantic incompleteness or overfitting to the oracle patch.




\phead{Results.}
After manual inspection, we find that all agent-generated patches killed by our generated tests are indeed invalid. We further identify three common types
of behavioral checks in the augmented tests that expose gaps in agent-generated patches, achieving Cohen’s $\kappa = 0.92$ .

\textit{\textbf{Input Selection for Behavioral Differentiation (36, 54.5\%).}} The augmented tests introduce inputs that differentiate correct behavior from plausible approximations. For example, in \texttt{django-14017}, the original tests confirm that valid expression types are accepted but do not check that invalid types are rejected. The augmented tests add cases with values that evaluate to \texttt{True} but should be rejected, covering both acceptance and rejection behavior. As another example, in \texttt{scikit-learn-14496}, the original inputs produce identical results under both truncation and rounding. The augmented tests introduce values for which the two operations yield different outputs, making the behavioral difference observable.

\textit{\textbf{End-to-End Behavioral Validation (27, 40.9\%).}}
The augmented tests extend validation from partial and local checks to end-to-end behavior across dependent code paths. For example, in \texttt{django-11276}, the original tests check only that \texttt{escape()} produces the updated entity format. The augmented tests instead assert on the final URL output after the data flows through both \texttt{escape()} and \texttt{urlize()}, catching patches that fix the local function but leave downstream behavior incorrect.  

\textit{\textbf{State Transition Validation (3, 4.6\%).}}
The augmented tests verify that the fix holds after state transitions, not just in the initial object state. For example, in \texttt{django-12965}, the original tests validate a deletion optimization only on a freshly created queryset. The augmented tests first evaluate the queryset and then perform the deletion, exposing patches that work on fresh objects but fail after prior state changes.

\begin{findings}
The augmented tests expose flaws in previously accepted agent-generated patches through three types of behavioral checks: input selection for behavioral differentiation (54.5\%), end-to-end behavioral validation (40.9\%), and state transition validation (4.6\%). These checks generalize across agents while introducing minimal overfitting to the oracle patch. 
\end{findings}
\section{Threats to Validity}
\label{sec:threats}

We discuss the main threats to internal, external, and construct validity below.

\phead{Internal Validity.}
Some surviving variants may be semantically equivalent to the oracle patch, which would overstate the prevalence of test gaps. We mitigate this through AST-level normalization, deduplication, and a validation stage that filters generated tests that fail to generalize across behavior-preserving variants. Residual equivalent variants may remain, but they would primarily make our test-weakness analysis more conservative.

Generated tests may overfit implementation details of the oracle patch. We address this via behavior-preserving transformations that discard tests sensitive to superficial artifacts, removing 23\% of candidate tests. Our RQ4 analysis finds only one ambiguous rejection across all affected instances, indicating that residual overfitting risk is low.

The benchmark may overlap with the LLM's training data, which could influence the candidate variants and tests generated by \tool. Since SWE-bench is a public benchmark, this threat 
applies broadly to LLM-based studies conducted on the same benchmark. \tool mitigates this risk by treating the LLM only as a candidate generator: retained tests must pass on the reference patch, fail on at least one surviving variant through execution, and pass robustness validation under behavior-preserving transformations. Thus, possible overlap may affect the candidate pool and generation yield, while retained tests still satisfy the same acceptance criteria.

The manual classifications in RQ1 and RQ4 involve subjective judgment. Two authors independently labeled all cases (Cohen's $\kappa = 0.94$ and 0.92). Disagreements were few, 3 of 77 in RQ1 and 3 of 66 in RQ4, and we resolved each in a single discussion by re-examining the case against the category definitions.

\phead{External Validity.}
Our evaluation is conducted on SWE-bench Verified (500 Python instances from 12 repositories) so our results may not generalize to other languages, benchmarks, or proprietary codebases. However, our approach is not inherently tied to SWE-bench Verified: it applies to any benchmark providing an oracle patch and executable tests. The mutation operators and validation pipeline are instantiated for Python, but extending them to other languages is feasible given the availability of 
 mutation testing tools that support the target language's features (e.g., PIT~\cite{PIT} or Major~\cite{Major} for Java, 
Mull~\cite{Mull} for C/C++). 
We evaluate patches from the top-10 leaderboard tools, covering both agent-based and LLM-based strategies, though this does not exhaust all repair approaches.


\phead{Construct Validity.}
PROBE uses the oracle patch as its primary correctness reference. If the oracle patch is not the only acceptable implementation, augmented tests may reject valid alternatives; our RQ4 analysis finds only one such case. We assess test quality using coverage and assertion characteristics, complemented by variant kill rate and downstream patch-assessment results. 
All LLM-based components in our implementation use GPT-5-mini, leaving the sensitivity of \tool to other model families and deployment settings unevaluated. Different models may generate variants and tests with different levels of plausibility, diversity, and effectiveness, which may alter both the inadequacy gaps discovered and the number of gaps successfully addressed. Our execution-based validation and robustness checks reduce reliance on the model's self-assessment when deciding whether to retain a generated test.
However, future studies should evaluate \tool with a broader range of cloud-based and local LLMs to assess the effect of model choice on its results.

\section{Conclusion}\label{sec:conclusion}

This paper presented \tool, a test augmentation framework that systematically identifies and closes behavioral gaps in benchmark regression test suites. \tool generates semantically modified variants of the oracle patch, treats the surviving variants as diagnostic signals of under-constrained behavior, and synthesizes targeted tests that make the missing constraints explicit. 
Our evaluation on SWE-bench Verified shows that test weakness is widespread: 77\% of instances admit at least one surviving variant. Guided by these variants, \tool generates 1,014 validated tests across 211 instances, increases patch-region line and branch coverage by 10.8\% and 9.5\%, and adds richer behavioral checks than the original developer-written tests. When used to re-evaluate the top-10 repair agents, the augmented suites reduce resolved rates by 4.2\%--9.0\%, showing that many previously accepted patches exploit benchmark test gaps rather than fully satisfying the intended repair semantics.
These results suggest that benchmark evaluation is not solely a solution-generation problem but also a test-strength problem. By turning implicit test gaps into explicit behavioral checks, \tool provides a practical path toward more reliable benchmark-based assessment in automated software engineering. 

\section{Data Availability Statement}\label{sec:data_availability}
We have made our research publicly available. All data, including the code and experimental results, can be found here:~\cite{probe2026}

\begin{acks}
We acknowledge the support of the Government of Canada’s New
Frontiers in Research Fund (NFRF), [NFRFE-2024-00612].
\end{acks}

\balance
\bibliographystyle{ACM-Reference-Format}
\bibliography{references}










\end{document}